# Strain-induced pseudomagnetic field and Landau levels in photonic structures

Mikael C. Rechtsman[1*], Julia M. Zeuner[2*], Andreas Tünnermann[2], Stefan Nolte[2], Mordechai Segev[1] and Alexander Szameit[2]

[1]Technion – Israel Institute of Technology, Haifa 32000, Israel
[2]Institute of Applied Physics, Friedrich-Schiller-Universität Jena, Max-Wien-Platz 1, 07743 Jena, Germany

mcr@technion.ac.il

Magnetic effects at optical frequencies are notoriously weak[1]. This is evidenced by the fact that the magnetic permeability of nearly all materials is unity in the optical frequency range, and that magneto-optical devices (such as Faraday isolators) must be large in order to allow for a sufficiently strong effect[2]. In graphene, however, it has been shown that inhomogeneous strains can induce 'pseudomagnetic fields' that behave very similarly to real fields[3]. Here, we show experimentally and theoretically that, by properly structuring a dielectric lattice, it is possible to induce a pseudomagnetic field at optical frequencies in a photonic lattice, where the propagation dynamics is equivalent to the evolution of an electronic wavepacket in graphene. To our knowledge, this is the first realization of a pseudomagnetic field in optics. The induced field gives rise to multiple photonic Landau levels (singularities in the density of states) separated by band gaps. We show experimentally and numerically that the gaps between these Landau levels give rise to transverse confinement of the optical modes. The use of strain allows for the exploration of magnetic effects in a non-resonant way that would be otherwise inaccessible in optics. Employing inhomogeneous strain to induce pseudomagnetism suggests the possibility that aperiodic photonic crystal structures can achieve

**greater field-enhancement and slow-light effects than periodic structures via the high density-of-states at Landau levels. Generalizing these concepts to other systems beyond optics, for example with matter waves in optical potentials, offers new intriguing physics that is fundamentally different from that in purely periodic structures.**

Magnetism in photonic structures has recently shown promise for a number of applications, despite its fundamentally weak nature at optical frequencies. For example, it has been shown (at microwave frequencies) that gyromagnetic photonic crystals have 'topologically protected' non-reciprocal edge states, meaning that radiation occupying these states is propagating without scattering and is exceedingly robust against disorder[4,5]. A related work[6] theoretically proposes that a photonic "topological insulator" structure can be realized using coupled resonator systems[7], for applications in robust delay lines. Optical metamaterials[8], in chiral or nonchiral[9] form, yield induced magnetism at optical frequencies. However, optical metamaterials are currently extremely lossy, because they incorporate metallic elements and are based on sharp resonances. In contradistinction with these, the structure we propose here – a strained honeycomb photonic lattice -- contains no metals and is non-resonant, and is therefore loss-free. It exhibits pseudomagnetic effects as a result of the inhomogeneous strain applied to it, which make it aperiodic. The pseudomagnetism, while not breaking time-reversal symmetry, leads to the generation of discrete Landau levels in the spatial spectrum, which are impossible in a periodic structure. The Landau levels are highly degenerate (and thus have very high density-of-states), while gaps form in between the Landau levels and lead to transverse spatial confinement of incident light.

More than a decade ago, Kane and Mele[3] showed that inhomogeneously strained graphene (a honeycomb lattice of carbon atoms) results in similar physical behavior as adding an external magnetic field. In particular, engineering a certain inhomogeneous strain corresponding to a constant magnetic field gives rise to Landau levels separated by gaps[10]. Since then, graphene physics has been realized in the optical domain in the form of a honeycomb photonic lattice, which exhibits a similar mathematical description to carbon-based graphene, but with additional effects unique to electromagnetic waves, such as conical diffraction and solitons[11–15]. In the present work, we demonstrate the existence of pseudomagnetism in such photonic lattices by virtue of the inhomogeneous strain.

A photonic lattice is composed of a periodic array of waveguides that are evanescently coupled to one another. Photonic lattices, which are arranged in a geometry similar to arrays of photonic crystal fibres[16], have been employed in exploring a number of fundamental wave transport phenomena, including discrete solitons[17–20], Anderson localization[21], and edge state properties[22,23]. Figure 1(a) depicts such a waveguide array in a honeycomb-lattice configuration, the exact geometry in which carbon atoms are arranged in graphene. As shown, the system is invariant in the z-direction, but arranged as a honeycomb in the transverse (x,y) plane. The propagation of monochromatic light through this waveguide array may be described by coupled-mode equations[20]:

$$i\partial_z \Psi_n = \sum_{<m>} c(|\mathbf{r}_{n,m}|)\Psi_m, \qquad (1)$$

where $\Psi_n$ is the amplitude of the mode in the $n^{th}$ waveguide, the summation is taken over all three nearest-neighbor waveguides to the $n^{th}$, and z is the propagation distance of the light within the photonic lattice. The coupling strength between the waveguides is described by the function $c(r) = c_0 e^{-(r-a)/l_0}$, where $a$ is the nearest-neighbor spacing and $l_0$ is the coupling decay length. Notice that Eq. (1) is equivalent to the Schrodinger equation in the tight-binding limit. Indeed, it is this analogy between paraxial photonic structures and quantum problems that gave rise to many experimental studies of fundamental issues, which are otherwise very difficult (sometimes inaccessible) in quantum world[20,24]. For simplicity, in Eq. (1) we assume that the waveguides are sufficiently distant from one another, such that it is only necessary to incorporate nearest-neighbor coupling. Furthermore, it is assumed that each of the waveguides supports only a single mode. Notice that here the strain makes the coupling vary from site to site, unlike the coupling in uniform lattices where $c(r)$ is simply a constant $c$[17,20]. Equation (1) is mathematically equivalent to the Schrodinger equation for graphene in the tight-binding limit, where the propagation distance, $z$ replaces time, $t$. Thus, light that propagates through the photonic lattice diffracts in the z-direction just as electrons in the p-orbitals of graphene evolve in time. The transverse wavefunction of the light that emerges from the lattice is therefore equivalent to a finite-time evolution of an electron wavepacket in graphene. As we show below, we demonstrate the presence of a strain-induced magnetic field using this "time" evolution of light in the photonic lattice.

Our honeycomb photonic lattice is fabricated using the direct femtosecond laser-writing technique[25] by locally increasing the refractive index to define the waveguides making up the photonic lattice within the volume of a fused silica sample. A microscope image of the input facet of the photonic lattice is displayed in Fig. 1(b). Using this fabrication technique, the waveguides are elliptical in nature, with horizontal and vertical diameters of 11$\mu m$ and 3$\mu m$, respectively, and the lattice constant is $a=14\mu m$. Specific details of the physical properties of the waveguides used in the experiment are given in the Supplementary Information section. In the experiments, we use laser light at vacuum wavelength of 633$nm$ and the fused silica sample has a background refractive index of 1.45. By making the substitution $\Psi_n = \psi_n e^{i\beta z}$ to Eq. (1), where $\beta$ is the propagation constant, we obtain an eigenvalue equation for the eigenmodes of the system: $-\beta\psi_n = \sum_{<m>} c(|\mathbf{r}_{n,m}|)\psi_m$. In Fig. 1(c) we show the band structure of the system, that is, $\beta$ vs. $(k_x, k_y)$, the transverse wavevector; since the honeycomb lattice has two members in each unit cell, there are two bands (two-dimensional surfaces in Fig. 1(c)). The band structure exhibits "Dirac cones": conical singularities in the band structure where the top and bottom bands intersect at a single point. Two of the six Dirac cones shown are mathematically unique, and the others are equivalent to each other through the periodicity of the band structure in $(k_x, k_y)$-space. This band structure of the propagation constant is similar to that of electrons in graphene. Indeed, these similarities gave rise to a variety of ideas that have been carried over from graphene physics to optics and vice-versa[14,26,12,27,28].

It has been shown for graphene that for wavepackets that lie near a Dirac point, straining the system is mathematically equivalent to introducing magnetic fields[3]. In particular, for a given position-dependent strain tensor $U(\mathbf{r})$, a vector potential,

$A(\mathbf{r}) = \pm(u_{xx} - u_{yy}, -2u_{xy})/2l_0$, is introduced, where $u_{xx}$, $u_{xy}$ and $u_{yy}$ are the elements of the two-dimensional strain tensor $U(\mathbf{r})$, and the sign depends on which Dirac point is in question. Note that the expressions for the pseudomagnetic field in graphene and a honeycomb photonic lattice are identical, if we choose units such that $\hbar/e = 1$. Recall that the magnetic field is given by $B(\mathbf{r}) = \nabla \times A(\mathbf{r})$. By choosing a particular form of the strain tensor $U(\mathbf{r})$, it was demonstrated in graphene[10] that a vector potential corresponding to a constant magnetic field could be achieved, along with a total absence of electric field. In this vein, if the waveguides in our honeycomb photonic lattice are transversely displaced from their original positions in the following way, then a constant magnetic field is achieved: $(u_r, u_\theta) = qr^2(\sin 3\theta, \cos 3\theta)$, where $u_r$ and $u_\theta$ are the radial and azimuthal displacements, $r$ is the distance from an arbitrary origin, $\theta$ is the azimuthal angle, and $q$ is a parameter corresponding to the strength of the strain. The result of this displacement is a vector potential $A(\mathbf{r}) = \pm 4q(y, -x)/l_0$, which in the symmetric gauge corresponds to a magnetic field of strength $B = 8q/l_0$. Note that the induced pseudomagnetic field does not break z-reversal invariance, just like a pseudomagnetic field in atomic graphene does not break time-reversal symmetry. In other words, the system would behave identically if the photonic lattice were reflected about the z=0 plane. This is because half of the Dirac cones exhibit pseudomagnetic fields in the +z direction, while the other half are in the –z direction[23]. This runs counter to the case of a two-dimensional electron gas with an applied magnetic field, where time-reversal symmetry is broken. In fact, breaking time-reversal symmetry in non-magnetic photonic media requires temporal modulation[29].

When a magnetic field is introduced in a wave equation, Landau levels form in the spectrum. The eigenstates of the system, instead of being spread out over a range of $\beta$, congregate at discrete, highly degenerate levels. In the two-dimensional electron gas, Landau levels are directly responsible for the observation of discrete steps in the Hall conductivity, which is known as the quantum Hall effect. In honeycomb photonic lattices, the pseudomagnetic field also separates the spectrum into Landau levels with $\beta = \pm \omega \sqrt{N}$, where $N = 0,1,2,...$, where $\omega = 3\sqrt{B}c_0 a/2$. The spacing of these Landau levels is unique to honeycomb structures, and is strongly affected by the conical dispersion at the vicinity of the Dirac cones. Otherwise, in two-dimensional electron gases with quadratic dispersion, the Landau levels are equally spaced and do not include the zeroth Landau level[30]. The effect of the strain is schematically depicted for an infinite honeycomb lattice in Fig. 2a and 2b. Specifically, Fig. 2a shows one of the Dirac cones in an unstrained honeycomb lattice. When the system is properly strained, the states within the Dirac cone split up into highly degenerate Landau levels, with band gaps lying in between them (Fig. 2b). Note that this is an accurate description of the band structure only in the Dirac regions. Outside the Dirac regions, the eigenstates form a continuous band and cannot be described by the pseudomagnetic field. Still, at the vicinity of the Dirac points the splitting effect is significant.

In order to numerically demonstrate the presence of the Landau levels, we plot the values of $\beta$ (i.e., the spatial spectrum) in the Dirac region for an unstrained and a strained lattice, in descending order, in Figs. 2c and 2d. Small sections of these finite lattices are shown schematically as insets in the figures. In both cases, we use a system with ~9600 waveguides, and for Fig. 2d we use a strain of corresponding to $q = 0.0015a^{-1}$ and $l_0 = a/5$. The unstrained system has only armchair-type edges, no zig-zag or bearded edges. In Fig. 2c, the spectrum of the unstrained is shown near $\beta/c_0 = 0$, the Dirac region where the Landau levels will emerge upon introducing the strain (blue curve). In Fig. 2d, we see that the strain clearly induces kinks in the spectrum that correspond to values of the propagation constant in which many eigenstates reside: these are the Landau levels; their propagation constants may be calculated, as described above, to be $\beta = \pm 0.37\sqrt{N}c_0$. Note that the numerically calculated positions of the Landau levels coincide exactly with the analytically predicted values (which assume that the states at the vicinity of the Dirac point obey the Dirac equation). The states in between the levels are localized solely on the edges of the lattice, i.e., these are strictly edge states that are not present in the limit of infinite system size. In the quantum Hall effect, it is these edge states that give rise to the Hall conductivity. Furthermore, the strain introduces edge states that are normally associated with the zig-zag and bearded edges[31] at $\beta/c_0 = 0$. This is a result of the fact that, while the unstrained lattice may have only armchair edges (which do not possess any edge states), in the strained lattice the armchair edge is deformed and therefore may take on some character of the other types of edge terminations, which have edge states associated with them. As a result, some of the $N = 0$ Landau level eigenstates have significant power residing on the armchair edges of the sample. The introduction of Landau levels provides a new mechanism, only achievable in aperiodic systems, to achieve very high degeneracies (as opposed to the van Hove mechanism typical of photonic crystals, and associated with band edges[32]). Therefore, if strain-induced Landau levels can be achieved in a photonic crystal setting (or example a photonic crystal slab on a silicon chip[33], or a three-dimensional photonic crystal[32]), the high density-of-states can be used for enhancement of spontaneous emission and nonlinear wave mixing processes via the Purcell effect[34].

We now proceed to present the direct observation of the strain-induced pseudomagnetic field in the photonic lattice. In order to induce the field, we write the waveguides at the transverse positions that correspond to the inhomogeneous strain described above. We have written a number of different arrays corresponding to increasing degree of strain, corresponding to $q = 0a^{-1}$ through $q = 0.021a^{-1}$, where again $a$ is the nearest-neighbor distance. If a piece of graphene were strained to this extent, it would yield a magnetic field of 5500T. The top row in Fig. 3 depicts optical microscope images of the input facets of our honeycomb photonic lattice, for increasing levels of strain.

We explore the properties of our strained honeycomb photonic lattices by using light from a Helium-Neon laser operating at a wavelength of *633nm*, incident on the input facet of the array, strongly focused on a single waveguide that resides at the edge of the array. As we will now show, probing the edge of the sample provides a full picture of the nature of the bulk spectrum, for two reasons: (1) it provides a straightforward way of accessing the Dirac region of the spectrum; and (2) the degree of localization on the edge is a clear indicator of whether (or not) the propagation constant of the excited spatial eigenmodes lies in a bulk band gap between Landau levels. The position of the excited waveguide is marked by the arrow in each of the panels in the upper row of Fig. 3. Since the input beam is focused on a single waveguide, the initial wavefunction is

$\Psi_n(z=0) = \delta_{nl}$, where the beam is incident upon the $l^{th}$ waveguide. Importantly, this input waveguide resides on the armchair edge of the honeycomb lattice, which has no localized states associated with it[31]. Since the unstrained honeycomb lattice does not have edge states on the armchair edge, light starting on that edge immediately spreads away from it into the bulk. This fact is shown in both the simulations and the experimental results depicted in the middle and bottom rows of Fig. 3. For this reason, the light launched at a waveguide on the armchair edge of the unstrained honeycomb lattice spreads throughout the lattice, as shown by the photographs taken at the output facet (left column, middle and bottom rows). For larger strains, projecting the initial state of the system (the specific initial beam we use) onto all the eigenstates of the system results in an overlap exclusively within the Dirac region, as we demonstrate in the Supplementary Information section. As the degree of strain, $q$, is increased, the light becomes more and more localized on the armchair edge, remaining largely within the same waveguide where it was launched (right column, middle and bottom rows). For sufficiently large strain, light is confined to the edge when the states excited are in a bulk band gap, and thus the light attempting to tunnel from the excited edge waveguide cannot penetrate into the bulk. This is strong evidence of the existence of the band gaps that lie in between the Landau levels. As was mentioned previously, the strained armchair edge does have edge states associated with it, all lying at propagation constant $\beta/c_0 = 0$[31], like the zig-zag and bearded edge states. Since the honeycomb edge states are necessarily degenerate[35], i.e., they all possess the same propagation constant, a wavepacket (beam) made up of a superposition of such states will not spread. That is, dispersive effects, which would be manifested here by spreading of the beam as it is propagating with z, are absent, because all the edge states of this strained honeycomb are degenerate, hence all acquire phase at the same rate as they propagate, avoiding dispersive effects altogether. Therefore, despite the fact that these edge states are degenerate with the zeroth Landau level and are resonantly coupled to bulk states, they do not couple into the bulk, and the light is therefore confined on the edge. In summary, the strain leads to confinement for two reasons: (1) the initial beam excites states in the gap between Landau levels, which are confined to the edge; and (2) the initial beams excite degenerate edge states in the zeroth Landau level which cannot cause diffraction due to their degeneracy. In the following section, we provide conclusive evidence that the excited eigenmodes reside within the band gaps induced by the Landau levels, at the center of the spectrum.

It is now essential to prove that the lack of tunneling from the waveguide residing on the armchair edge into the bulk indeed arises from the presence of the gaps between Landau levels, and not from some coincidental perturbation near the edge which would constitute a simple "defect mode". To show that, we fabricate several different samples, all with a given strain value $q=0.021a^{-1}$, with that particular waveguide replaced by a "defect waveguide." Specifically, we change the refractive index of that waveguide (while leaving all others fixed), such that it is higher or lower than the others. When a defect mode is introduced, Eq. (1) must be changed to: $i\partial_z \Psi_n + \beta_l \delta_{nl} \Psi_n = \sum_{<n,m>} c(|\mathbf{r}_{nm}|) \Psi_m$, where $\beta_l$ is the intrinsic propagation constant of the defect waveguide mode. The dimensionless parameter $\beta_l/c_0$ describes the strength of the defect mode: it is positive if the change in refractive index is positive and negative if the change in refractive index is negative. Details of the calculation of $\beta_l/c_0$ for a defect waveguide of given propagation constant are presented in the Supplementary Information section. We schematically depict the nature of the entire band structure of the strained honeycomb lattice in Fig. 4(a): in the center (near the Dirac-cone regions) lie the Landau levels (symmetric on either side of $\beta = 0$); on either side of the Landau levels lie the eigenstates of the photonic lattice which are outside the Dirac region. These regions are largely composed of bulk eigenstates that penetrate throughout the lattice. In Fig. 4(b), we show the output facet of the lattice where there is no defect ($\beta_l/c_0 = 0$): light is tightly confined because it has excited localized eigenstates lying within Landau-level gaps that are localized near the edge of the lattice. Upon slightly decreasing the refractive index of the defect waveguide (such that $\beta_l/c_0 = -0.5$), shown in Fig. 4, the beam at the output facet expands slightly, reflecting increasing overlap with the continuum modes below the Landau levels in the band structure. Upon significantly decreasing the refractive index of the waveguide ($\beta_l/c_0 = -2$), the excited eigenstates are bulk states, well away from the Landau levels and thus, light significantly spreads into the bulk of the structure (see Fig. 4(d)). At this point, we consider defect waveguides with refractive index larger than the ambient index of the rest of the waveguides. Just as in the negative-defect case, a slight increase in the refractive index, corresponding to $\beta_l/c_0 = 0.5$, leads to a slight increase in the expansion of the beam by the time it gets to the output facet of the lattice (see Fig. 4(e)). However, a much stronger increase, corresponding to $\beta_l/c_0 = 2$, shows a dramatic expansion of the beam because the excited eigenstates (which are spectrally distant from the Dirac region where the Landau levels reside) now lie throughout the bulk of the photonic lattice (see Fig. 4(f)). A further increase in the refractive index (such that $\beta_l/c_0 = 4$) results in relocalization of the beam (see Fig. 4(g)). The reason that the beam localizes again for sufficiently strong refractive index is that the defect eigenstate has passed through the entire band and now lies above it, acting much like a localized donor or acceptor mode in a semiconductor. Outside the band, the state resides in a band gap (although not one induced by the strain), and thus cannot overlap with any modes that spread throughout the bulk of the photonic lattice. Taken together, the fact that the negative and positive defects both lead to spreading proves conclusively that the gap, in between the strain-induced Landau levels, lies at the center of the band.

The introduction of strong magnetic effects in optical systems opens the door to a wide range of new physical effects and applications. For example, recent work on gyrotropic photonic crystals[5] has shown in the microwave regime that quantum-Hall physics may be used to achieve scatter-free propagation in optical isolators. The fact that the pseudomagnetic field has the effect of "grouping" eigenstates together into highly degenerate Landau levels suggests the possibility of using this effect for extreme efficiency enhancement in nonlinear devices in photonic crystals. Fundamental questions remain to be answered, for example: can a pseudomagnetic field be observed using strain in a photonic crystal slab geometry (i.e., a two-dimensional geometry with a finite height in the third direction), in order to utilize the field on a silicon chip? Can this high density-of-states be engineered by strain in a fully three-dimensional photonic crystal? What is the nonlinear enhancement associated with Landau levels? The strained photonic lattice provides an excellent experimental setting for probing both linear and nonlinear effects of magnetism in optics. Furthermore, the recent observation of parity-time-symmetry (PT) breaking in optics[36,37] has very intriguing implications in honeycomb PT-symmetric lattices[13,38], which suggests that the strained honeycomb lattice may provide a context for understanding the effect of magnetism on the PT transition, and on non-Hermitian optics in general[39–41]. Can terahertz generation be enhanced in photonic crystals via strain? What is the nature of wave mixing processes between highly degenerate Landau levels? Can lasing thresholds in photonic crystal cavities be further reduced? In the plethora of applications where structured photonics have an important technological impact, a new important question will be: can we make them better via an inhomogeneous strain?

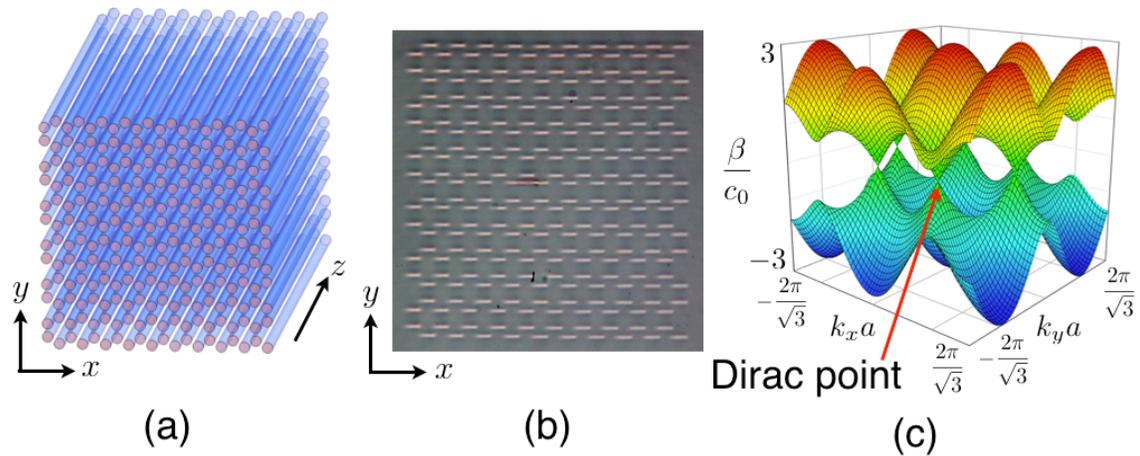

Figure 1. (a) Diagram of the honeycomb photonic lattice geometry. Light propagates through the structure along the axis of the waveguides (the z-axis) through tunneling between neighboring waveguides. (b) Microscope image of the input facet of the photonic lattice geometry. The waveguides are elliptical (due to fabrication constraints), with dimensions of $11\mu m$ in the horizontal direction and $3\mu m$ in the vertical direction. (c) Band structure diagram of the photonic lattice, with $\beta/c_0$ plotted as a function of the

Bloch wavevector $(k_x, k_y)$. Note that the first and second bands intersect at the Dirac cones (one of which is indicated by an arrow), which reside at the vertices of the Brillouin zone.

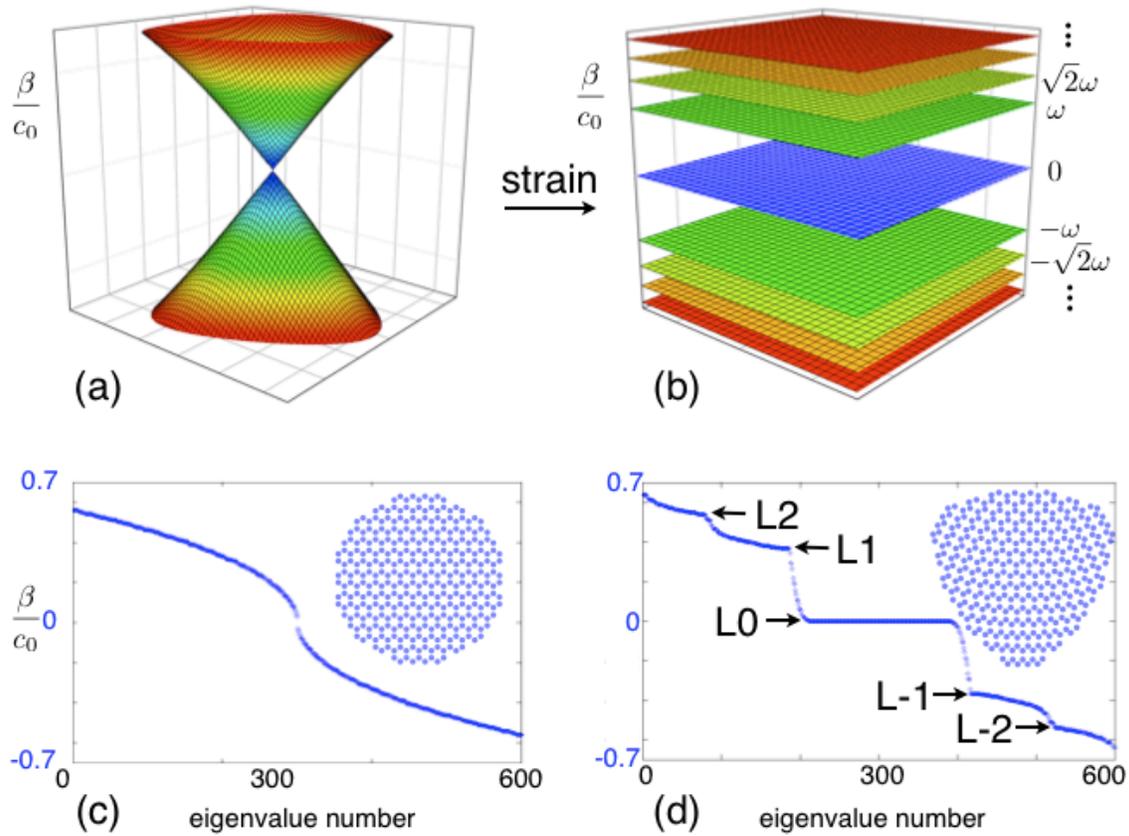

Figure 2. (a) Schematic depiction of a Dirac cone in the spectrum of the unstrained honeycomb photonic lattice. (b) The Dirac cone splits into Landau levels, with $\frac{\beta}{c_0} = \pm\sqrt{N}$ where N=0,1,2,…, upon straining the system as described in the text. (c) The numerically-computed eigenvalues plotted in ascending order in the region near the Dirac

point for the unstrained lattice. Inset: a circular section of the lattice. (d) Numerically-computed eigenvalues for the strained system in the Dirac region, as specified in the text. Inset: the effect of the strain on the section of the honeycomb lattice shown in (c). Clear Landau levels emerge in the spectrum as a result of the strain (labeled 'LL'), with edge states lying in between them. The calculations for (c) and (d) employ 9600 waveguides. The strain in (d) is given by $q = 0.0015a^{-1}$.

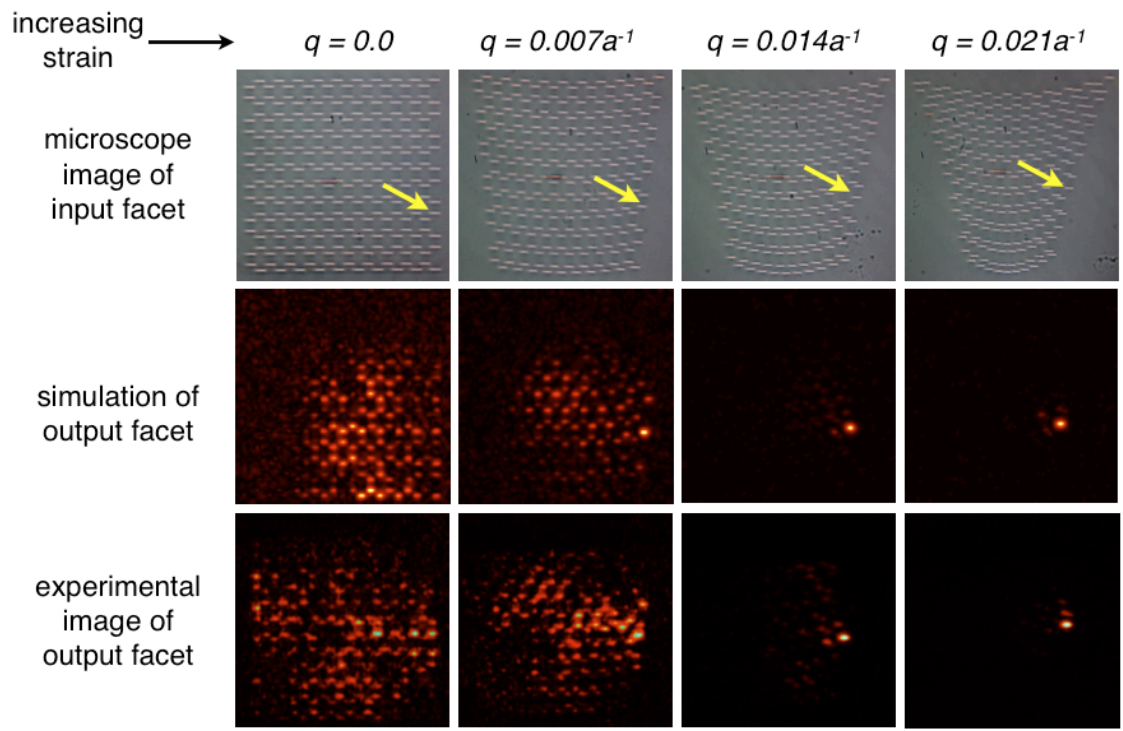

Figure 3. Experimental and simulation results for increasing strain. Top row: microscope images of the input facet of the lattice. The yellow arrow marks the waveguide into which the input light beam is launched. This waveguide resides at the armchair edge, which has no edge states. Center row: simulation results showing the intensity profile at the output

facet when light is launched into the edge waveguided (marked by the arrow in the upper row). Bottom row: experimental results showing the intensity profiles of the light exiting the lattice, for light launched into the edge waveguide marked by the yellow arrow in (a). Results for the unstrained system ($q=0.0a^{-1}$) are shown in the left-most column. Due to the lack of edge states on the armchair edge, no light is confined to the edge and it diffracts into the bulk. The second, third and fourth columns are similar but with successively increasing strains ($q = 0.007a^{-1}$, $0.014a^{-1}$, and $0.021a^{-1}$). Light becomes highly confined on the edge with increasing strain. As described in the text, this results from the fact that the eigenstates excited by the incoming light are either in a band gap between Landau levels (and therefore cannot penetrate into the bulk), or are degenerate states in the zeroth Landau level, also resulting in edge confinement.

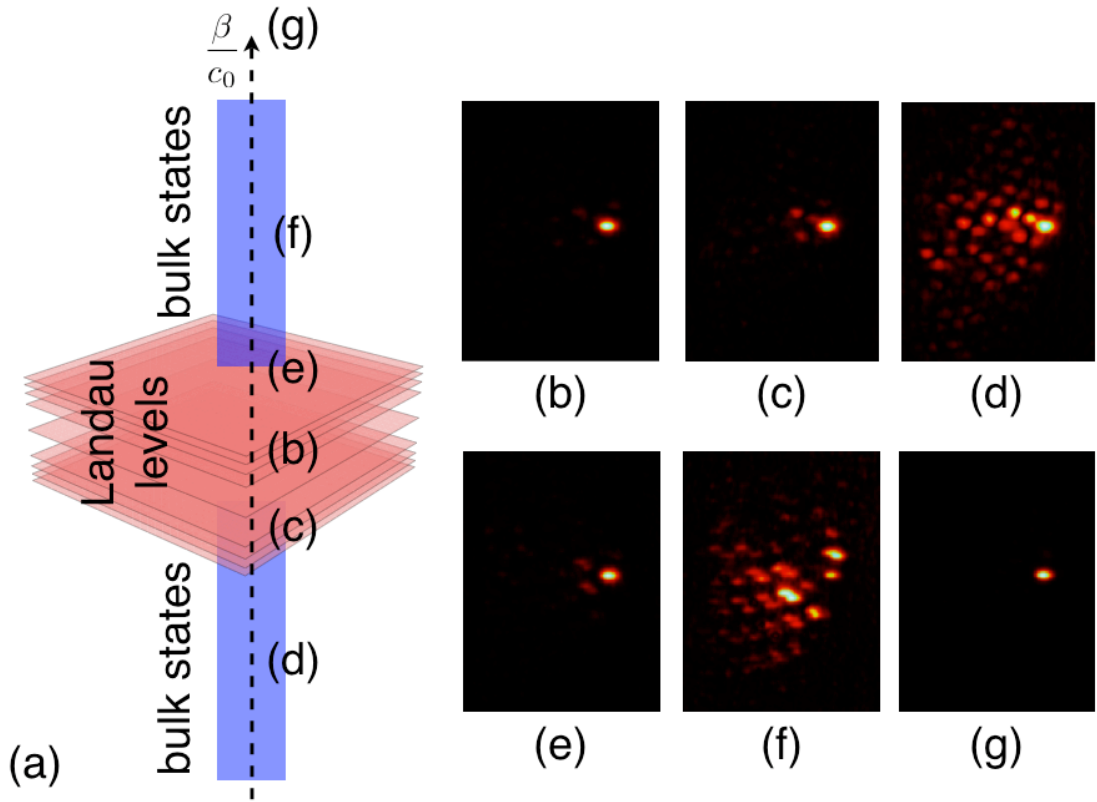

Figure 4. (a) Schematic depiction of the eigenvalue spectrum of a strained honeycomb photonic lattice. The Landau levels reside at the center of the spectrum, where the Dirac cones lie in the unstrained lattice. (b) Experimental results showing the light exiting the properly strained honeycomb photonic lattice, when a single waveguide on the armchair edge is excited. The light is confined as a result of the gaps between the Landau levels, as described in the text and as shown in Fig. 3. (c) Light exiting the strained lattice when the excited waveguide is engineered to have a slightly negative defect ($\beta_l/c_0 = -0.5$). As a result, coupling with the bulk bands causes some spreading. (d) Larger defect strengths ($\beta_l/c_0 = -2$) causes more spreading of the light into the photonic lattice. (e) A mildly positive defect ($\beta_l/c_0 = 0.5$) also causes some spreading, as in (c). (f) A stronger positive defect ($\beta_l/c_0 = 2$) causes greater spreading (as in (d)) as a result of strong

coupling to the bulk bands. (g) Once the defect mode is very strong ($\beta_l/c_0 = 4$) it becomes a defect mode outside of the band and therefore in another band gap. As a result, it becomes localized again. These results, taken together, situate the excitation of the defect-free case ($\beta_l/c_0 = 0$) within the Landau level gaps.

# *Supplementary Information*

# Strain-induced pseudo-magnetic field and Landau levels in photonic structures

Mikael C. Rechtsman[1*], Julia M. Zeuner[2*], Andreas Tünnermann[2], Stefan Nolte[2], Mordechai Segev[1] and Alexander Szameit[2]

[1]Technion – Israel Institute of Technology, Haifa 32000, Israel
[2]Institute of Applied Physics, Friedrich-Schiller-Universität Jena, Max-Wien-Platz 1, 07743 Jena, Germany

*mcr@technion.ac.il*

In this Supplementary Information section, we provide specific details on the structure of the femtosecond-written waveguides, the strained honeycomb photonic lattice, and the rationale behind exciting the lattice at the edge in order to probe the Landau-level wave dynamics.

**Section 1: Femtosecond-written waveguides and their properties**

As described in the text body, we employ direct femtosecond-written waveguides arranged in a strained honeycomb lattice in order to explore the effects of the pseudomagnetic field on transverse wave propagation. There, we describe the wave dynamics in dimensionless terms in terms of coupled mode theory; in the present section we specify the physical parameters associated with this theory.

In Fig. S1, we show the structure of a single waveguide. It is written using a pulsed Titanium:Sapphire laser operating at a wavelength of *800nm*. It is elliptical because the focus of the pulsed laser beam is broader in the longitudinal direction than the transverse direction. Its dielectric structure can be described as a hypergaussian function:

$$\Delta n_w(x,y) = \Delta n_0 \exp\left(-\left[\left(\frac{x}{r_1}\right)^2 + \left(\frac{y}{r_2}\right)^2\right]^3\right). \qquad (S1)$$

Here, $\Delta n_w(x,y)$ is the refractive index change associated with the waveguide. We use fused silica as the background medium, with a refractive index of *$n_0$=1.45*, $\Delta n_0$ is the maximum refractive index change within the waveguide, and $r_1$ and $r_2$ are the radii in the horizontal and vertical directions respectively. For the waveguides used in the present work, we have $\Delta n_0 = 7\times 10^{-4}$, $r_1 = 5.5\mu m$ and $r_2 = 1.9\mu m$.

## Section 2: Calculating guided eigenmodes of femtosecond-written waveguides

The equation describing the propagation (i.e., diffraction) of light through the array of waveguides is the Schrödinger-like paraxial wave equation:

$$i\partial_z \psi(x,y,z) = -\frac{1}{2k_0}\nabla^2 \psi(x,y,z) - \frac{k_0 \Delta n(x,y)}{n_0}\psi(x,y,z), \quad (S2)$$

where $z$ is the distance of propagation along the propagation axis of the photonic lattice; $\psi$ is the envelope function of the electric field as defined by $E(x,y,z) = \psi(x,y,z)e^{i(k_0 z - \omega t)}\hat{x}$ (where $E$ is the electric field, $k_0$ is the wavenumber within the medium and $\omega = ck_0/n_0$); $\Delta n(x,y)$ is the profile of the refractive index, and $\nabla^2$ is the Laplacian in the (x,y) plane. In the present work, $\lambda = 633 nm$ (the wavelength of a Helium-Neon laser).

In order to solve for the guided modes of this waveguide, we employ a plane-wave expansion method[1] applied to the Schrödinger equation, and diagonalize the equation in the MATLAB simulation software. This yields both the guided mode of the waveguide (in the present work the waveguides are designed to be single-moded), as well as the propagation constant of the mode (which is the eigenvalue of the operator defined on the left-hand side of Eq. (S2)). In Fig. S1 we show the structure of an eigenmode of a single waveguide, as calculated using the plane-wave expansion method.

The basis of our analysis of the strained honeycomb structure introduced in this paper is coupled-mode (also referred to as tight-binding) theory, as presented in Eq. (1). In coupled-mode theory, we assume that the continuum wavefunction $\psi(x,y,z)$ can be written as the sum of waveguide modes centered on each waveguide. Thus, the full infinite-dimensional dynamics of the Schrödinger equation can be reduced to a discrete set of equations involving only the amplitudes and phases of the waveguide modes (given in Eq. (1)). In order to calculate the physical coupling constant, $c_0$, between waveguides in the unstrained system, we simply calculate the eigenvalues of a system with two waveguides at the nearest-neighbor distance of $14 \mu m$. The splitting in the eigenvalues should be exactly twice the coupling constant[2]. In this way, we obtain a value of $c_0 = 1.9 \times 10^{-4} \mu m^{-1}$. Note that despite the anisotropy of the waveguide itself, the mode associated with the waveguide is quite isotropic and so the anisotropy of the coupling is negligible. Furthermore, we have confirmed that next-nearest-neighbor hopping terms are negligible compared to nearest-neighbor terms.

The discussion above explains the relationship between the refractive index of a waveguide and its relative propagation constant. To complete the discussion, let us now consider a defect waveguide, in relation to Fig. 4 in the paper, where we demonstrate the effect of adding a defect waveguide in the pseudomagnetically-strained honeycomb lattice. As we show in that figure, if the defect waveguide has a refractive index that is either slightly higher or slightly lower than the rest of the waveguides in the array, this leads to spreading into the bulk. Namely, unlike what is happening in an unstrained lattice – where light launched into a defect waveguide remains in that waveguide throughout propagation, here – in the presence of the pseudomanetic field – the light escapes from the defect waveguide. In the paper, we present for a number of different realizations of the defect, the value of its propagation constant relative to the coupling constant, $\beta_l/c_0$. The procedure described in this section explains how this quantity may be calculated for a given defect waveguide.

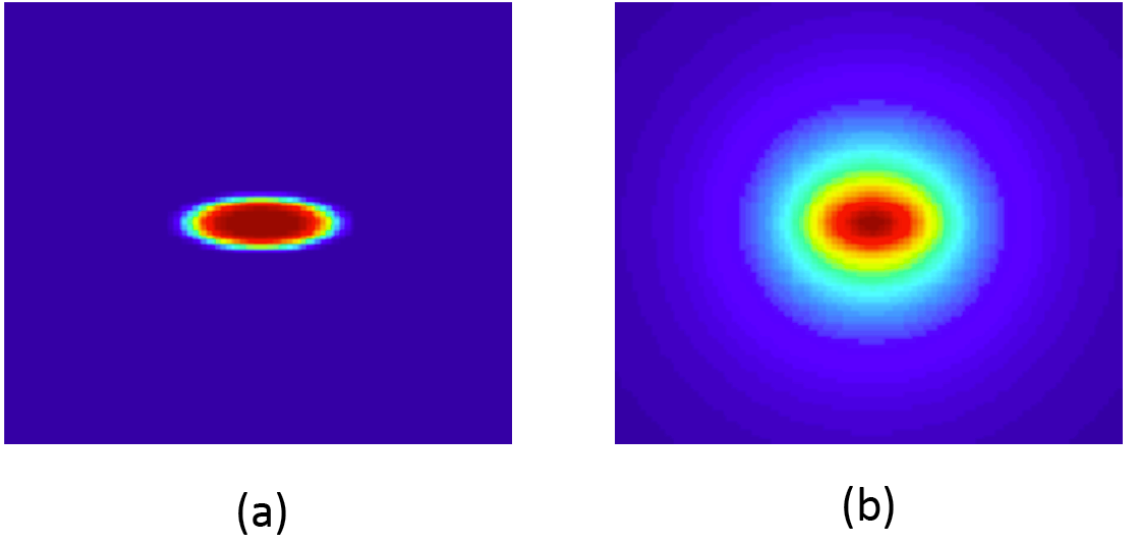

FIG S1: Waveguide and guided mode. (a) shows the refractive index profile of the waveguide, and (b) shows the field-strength profile of the guided eigenmode.

**Section 3: Coupled-mode theory analysis: exciting single waveguides on armchair edge excites modes entirely within the Dirac region**

In the paper, we claim that by launching light in a single waveguide on the armchair edge of the strained photonic lattice, we only excite modes within the Dirac region. In this way we can directly probe the effects of the pseudomagnetic field on the structure. Furthermore, the light that is incident on the armchair edge excites modes that are within gaps between Landau levels, and those in the zeroth level. Notably, the modes in the zeroth Landau level result from distortions of the armchair edge, and their features resemble those of the edge states of the zig-zag or bearded edges. In the present section, we demonstrate (using coupled mode theory) that this is in fact the case.

Consider the calculation depicted in Fig. 2 of the paper. Here, a tight-binding model is used to calculate the spatial spectrum (values of $\beta$) in the vicinity of the Dirac region (around $\beta = 0$ between $\beta = -0.6c_0$ to $\beta = 0.6c_0$). The unstrained lattice consists of only armchair edges, and therefore has no edge states associated with it whatsoever. In Fig. S2, we describe the projection of a typical single waveguide on the armchair edge (in other words, the wavefunction $\psi_n = \delta_{m,n}$, where the $m^{th}$ waveguide resides on the armchair edge). In the figure, we plot the projection $|\langle\psi|\phi_l\rangle|$, where $|\phi_l\rangle$ is the $l^{th}$ eigenstate of the system, for both the unstrained system and the strained system. In the unstrained system, the total fraction of power that resides in eigenstates in the Dirac region is 9%. Indeed, without the strain, the power is spread quite evenly over all eigenmodes, and the relatively low amount of power in the Dirac region is a result of the relatively low density-of-states there. However, in the strained system (with a strain given by $c=0.0015a^{-1}$, the same as that considered in Fig. 2), 55% of the power resides in the Dirac region. We find that for increasing strain, a monotonically increasing fraction of the projected power resides in the Dirac region. Thus, exciting the armchair edge for large values of the strain entirely confines light to being in the Dirac region of the spectrum. Also depicted in Fig. S2 are typical wavefunctions associated with a strained system, such as a typical first Landau level, and an edge state in between Landau levels. It is clear from Fig. S2 that the power projected on the eigenmodes resides in modes in between Landau levels and to some extent in the zeroth Landau level, as described in the paper.

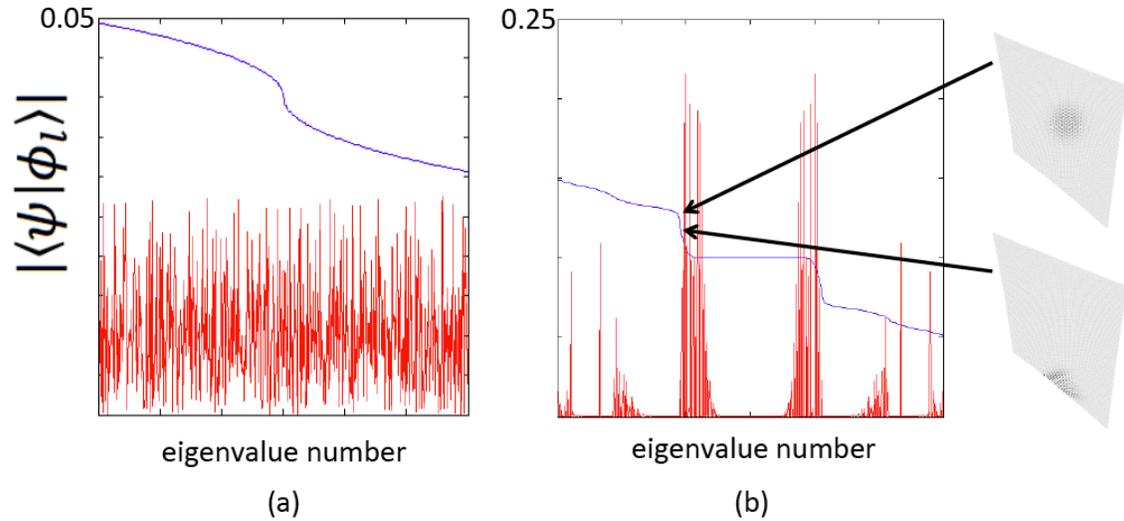

Figure S2: Projection of a single-waveguide excitation on the armchair edge onto the eigenstates of the system, $|\langle\psi|\phi_i\rangle|$, within the Dirac region ($\beta \in [-0.6c_0, 0.6c_0]$) (red curves). The spatial spectrum of the eigenmodes is shown in blue as a guide to the eye of where the Landau levels (as well as the edge states in between them) reside.